# Crushing of interstellar gas clouds in supernova remnants. I. The role of thermal conduction and radiative losses

S. Orlando[1], G. Peres[2], F. Reale[2], F. Bocchino[1], R. Rosner[3,4], T. Plewa[3,4], and A. Siegel[3,4]

[1] INAF - Osservatorio Astronomico di Palermo "G.S. Vaiana", Piazza del Parlamento 1, 90134 Palermo, Italy
[2] Dip. di Scienze Fisiche & Astronomiche, Univ. di Palermo, Piazza del Parlamento 1, 90134 Palermo, Italy
[3] Dept. of Astronomy and Astrophysics, University of Chicago, 5640 S. Ellis Avenue, Chicago, IL 60637, USA
[4] Center for Astrophysical Thermonuclear Flashes, University of Chicago, 5640 S. Ellis Avenue, Chicago, IL 60637, USA

**Abstract** We model the hydrodynamic interaction of a shock wave of an evolved supernova remnant with a small interstellar gas cloud like the ones observed in the Cygnus loop and in the Vela SNR. We investigate the interplay between radiative cooling and thermal conduction during cloud evolution and their effect on the mass and energy exchange between the cloud and the surrounding medium. Through the study of two cases characterized by different Mach numbers of the primary shock ($\mathcal{M} = 30$ and 50, corresponding to a post-shock temperature $T \approx 1.7 \times 10^6$ K and $\approx 4.7 \times 10^6$ K, respectively), we explore two very different physical regimes: for $\mathcal{M} = 30$, the radiative losses dominate the evolution of the shocked cloud which fragments into cold, dense, and compact filaments surrounded by a hot corona which is ablated by the thermal conduction; instead, for $\mathcal{M} = 50$, the thermal conduction dominates the evolution of the shocked cloud, which evaporates in a few dynamical time-scales. In both cases we find that the thermal conduction is very effective in suppressing the hydrodynamic instabilities that would develop at the cloud boundaries.

## 1. Introduction

One of the primary reasons for the great morphological complexity of the shells of supernova remnants (SNRs) is the inhomogeneity of the interstellar medium (ISM) where they propagate (e.g. Hester 1987). Optical, UV and X-ray observations of SNRs show that the SN-generated shock waves travel through and interact with the denser clouds they encounter (e.g. Bocchino et al. 2000; Levenson et al. 2002; Patnaude et al. 2002; Nichols & Slavin 2004; Levenson & Graham 2005; Miceli et al. 2005), generating transmitted and reflected (bow) shocks, which, in turn, interact with each other (e.g. McKee & Cowie 1975; White & Long 1991; Poludnenko et al. 2002). Knowledge of how the SNR shocks interact with the inhomogeneous ISM and interstellar clouds is very important for the study of the interstellar gas dynamics itself, for our understanding of the emission of this process, and for the detailed analysis of the mass distribution of the plasma in the Galaxy, energy exchanges, and the chemical composition and evolution of the ISM.

The overall evolution of the shock-cloud interaction has been studied analytically by a number of authors (e.g. McKee & Cowie 1975; Heathcote & Brand 1983; McKee et al. 1987). However, the complex nature of the phenomenon, involving the

*Send offprint requests to*: S. Orlando,
e-mail: orlando@oapa.astropa.unipa.it

non-linear interaction among thermally conducting supersonic flows, radiative losses, magnetic fields, non-equilibrium ionization effects, etc., and the comprehension of the details of the mass and energy exchange between the cloud and the intercloud medium have required numerical simulations.

The first numerical studies of this problem showed that hydrodynamic Kelvin-Helmholtz (KH) and Rayleigh-Taylor (RT) instabilities develop at the interface between the cloud and the post-shock ambient medium due to the shear motions around the cloud and to the density gradients at the cloud boundary (Woodward 1976; Bedogni & Woodward 1990). Klein et al. (1994) (hereafter KMC94) studied extensively the interaction of a strong shock with a single, non-radiative cloud with 2-D calculations. They explored the parameter space defined by the Mach number $\mathcal{M}$ and the initial density contrast cloud/surrounding medium $\chi$, and showed that the cloud is ultimately destroyed within a few dynamical time-scales by the hydrodynamic instabilities (see also Poludnenko et al. 2002 for the interaction of a strong shock with a system of clouds).

The first 3-D calculations of the shock-cloud interaction showed a richer structure of the hydrodynamic instabilities (Stone & Norman 1992; Xu & Stone 1995). The 3-D calculations do not invalidate any of the conclusions drawn from the 2-D calculations; however, they showed that the cloud breaks in all directions and that the turbulent mixing of the cloud and the interstellar medium is complete, with the formation of macro-scopic vortex filaments. The details of the cloud mass mixing, therefore, may be different if computed in 2- or 3-D.

The interaction of the shock with a *radiative* cloud has been only recently analyzed in detail (Mellema et al. 2002; Fragile et al. 2004). 2-D calculations have shown large differences from the non-radiative case: the compressed radiating cloud breaks up into numerous dense and cold fragments. The cooling processes can be highly efficient already for moderate cloud densities ($\gtrsim 1$ cm$^{-3}$) and shock Mach number ($\lesssim 20$). 2-D calculations of the interaction between magnetized shocks and radiative clouds (Fragile et al. 2005) showed that the magnetic field may increase the efficiency of the radiative cooling, acting as a confinement mechanism ultimately driven by the interstellar flow and local field stretching.

The interaction of a strong shock with a *thermally conducting and radiative* cloud has been less studied so far. Vieser & Hensler (2000) and Hensler & Vieser (2002) investigated the effect of the heat conduction in the context, different from that of this paper, of a giant self-gravitating cloud moving subsonically (i.e. in the absence of shock waves) through an hot diluted medium; they showed that the thermal conduction suppresses the hydrodynamic instabilities leading to cloud disruption, so that the cloud is stabilized and survives. However, a detailed and systematic analysis of the competition of the radiative losses and the thermal conduction in the evolution and in the energy exchanges of the shock-cloud system is still missing. Nevertheless, this competition may be important in determining both the local and the global dynamics of the shocked cloud. In particular, the radiative losses may induce thermal instabilities, and lead to the cloud fragmentation in cold and dense cloudlets (Mellema et al. 2002; Fragile et al. 2004). The thermal conduction may instead act to save the cloud (KMC94) and, therefore, to reduce the mixing of cloud material with the ambient medium, through the suppression of the hydrodynamic instabilities. In addition, the thermal conduction leads to the heating and evaporation of the shocked cloud, reducing the radiative cooling.

In spite of the extensive literature on this subject, several aspects of the shock-cloud interaction remain unexplored: How do the effects of radiative losses and thermal conduction combine on the interaction and subsequent evolution of unmagnetized shocked clouds? How and under which physical conditions can the magnetic fields suppress the thermal conduction during the cloud evolution? To what extent do the complex dynamics of the shock-cloud interaction induce deviations from the equilibrium of ionization of the cloud medium? What observational features (e.g. morphological, spectral) in the optical, UV, and X-ray bands are tracers of the physical processes at work and can, thus, provide accurate diagnostics of shocked cloud plasma?

To answer these questions, we have started a new project devoted to study the shock-cloud interaction through detailed and extensive numerical modeling. The project aims at overcoming some of the limitations found in the previous analogous studies and crucial for the accurate interpretation of the high resolution multi-wavelength observations of middle-aged SNR shell available with the last-generation instruments. In the present paper, the detailed numerical study focuses specifically on the interplay between the radiative losses and the thermal conduction, the latter including both the classical and the free-streaming regime (Cowie & McKee 1977). Our approach is to consider two examples of shock-cloud interaction, with different shock Mach numbers, as test cases in which one or the other of the two processes plays a dominant role. For each of these cases, we identify the effects of the thermal conduction and of the radiative losses on the dynamics, by comparing models calculated with these physical processes turned "on" or "off". We perform 2-D hydrodynamic simulations and also 3-D simulations, where necessary.

In future papers, we plan to analyze the deviations from the equilibrium of ionization during the shock-cloud dynamics and the observable properties of the shocked clouds, including spectra synthesized at different evolutionary phases.

The paper is organized as follows: Sect. 2 describes the numerical setup and the characteristic physical parameters of the problem; in Sect. 3 we investigate the role played by the thermal conduction and by the radiative losses in the dynamics of the shocked cloud and how such a role changes under different physical conditions; in Sect. 4 we summarize the results and draw our conclusions.

## 2. Problem description and numerical scheme

We study the impact of a shock wave on an isolated unmagnetized spherical cloud. The cloud is assumed to be small compared to the curvature radius of the shock. These assumptions and the adopted parameters of the shock wave are characteristic of a SNR forward shock overruning a small interstellar cloud[1]. We assume that the fluid is fully ionized, and may be regarded as a perfect gas (with a ratio of specific heats $\gamma = 5/3$).

### 2.1. The model equations

We consider a numerical model based on the solution of the Euler equations, taking into account the effects of the radiative losses from an optically thin plasma and of the thermal conduction (including the effects of heat flux saturation):

$$\frac{\partial \rho}{\partial t} + \nabla \cdot \rho \mathbf{u} = 0 ,$$

$$\frac{\partial \rho \mathbf{u}}{\partial t} + \nabla \cdot \rho \mathbf{u}\mathbf{u} + \nabla P = 0 , \quad (1)$$

$$\frac{\partial \rho E}{\partial t} + \nabla \cdot (\rho E + P)\mathbf{u} = -\nabla \cdot q - n_{\mathrm{e}} n_{\mathrm{H}} \Lambda(T) .$$

Here $\quad E = \epsilon + \frac{1}{2}|\mathbf{u}|^2 ,$

is the total gas energy (internal energy, $\epsilon$, and kinetic energy), $t$ is the time, $\rho = \mu m_H n_{\mathrm{H}}$ is the mass density, $\mu = 1.26$ is the mean atomic mass (assuming cosmic abundances), $m_H$ is the mass of the hydrogen atom, $n_{\mathrm{H}}$ is the hydrogen number density, $n_{\mathrm{e}}$ is the electron number density, $\mathbf{u}$ is the gas velocity, $T$ is the temperature, $q$ is the conductive flux, and $\Lambda(T)$ represents the radiative losses per unit emission measure[2] (e.g. Raymond & Smith 1977; Mewe et al. 1985; Kaastra & Mewe 2000). We use the ideal gas law, $P = (\gamma - 1)\rho\epsilon$.

---

[1] For typical galactic SNR, such conditions are met during the Sedov-Taylor expansion phase (KMC94).

To allow for a smooth transition between the classical and saturated conduction regime, we followed Dalton & Balbus (1993) and defined the conductive flux as

$$q = \left(\frac{1}{q_{\rm spi}} + \frac{1}{q_{\rm sat}}\right)^{-1}. \tag{2}$$

Here $q_{\rm spi}$ represents the classical conductive flux (Spitzer 1962)

$$q_{\rm spi} = -\kappa(T)\nabla T \tag{3}$$

where the thermal conductivity is $\kappa(T) = 5.6 \times 10^{-7} T^{5/2}$ erg s$^{-1}$ K$^{-1}$ cm$^{-1}$. The saturated flux, $q_{\rm sat}$, is (Cowie & McKee 1977)

$$q_{\rm sat} = -{\rm sign}\,(\nabla T)\,5\phi\rho c_{\rm s}^3, \tag{4}$$

where $c_{\rm s}$ is the isothermal sound speed, and $\phi$ is a number of the order of unity. We set $\phi = 0.3$ according to the values suggested for a fully ionized cosmic gas: $0.24 < \phi < 0.35$ (Giuliani 1984; Borkowski et al. 1989, Fadeyev et al. 2002, and references therein); we assumed that electron and ion temperatures are equal[3].

In order to trace the motion of the cloud material, we consider a passive tracer associated with the cloud. To this end, we add the equation

$$\frac{\partial C_{\rm cl}}{\partial t} + \nabla \cdot C_{\rm cl}\mathbf{u} = 0 \tag{5}$$

to the standard set of hydrodynamic equations. $C_{\rm cl}$ is the mass fraction of the cloud inside the computational cell. The cloud material is initialized with $C_{\rm cl} = 1$, while $C_{\rm cl} = 0$ in the ambient medium. During the shock-cloud evolution, the cloud and the ambient medium mix together, leading to regions with $0 < C_{\rm cl} < 1$. At any time $t$ the density of cloud material in a fluid cell is given by $\rho_{\rm cl} = \rho C_{\rm cl}$.

The calculations described in this paper were performed using the FLASH code (Fryxell et al. 2000). FLASH is an adaptive mesh refinement multiphysics code. For the present application, the code has been extended by additional computational modules to handle the radiative losses and the thermal conduction in the formulation of Spitzer (1962). The hydrodynamic equations are solved using the FLASH implementation of the PPM algorithm (Colella & Woodward 1984). The thermal conduction is treated using an operator-splitting method with respect to the hydrodynamic evolution: the heat flux is calculated with a finite difference explicit formula and then added to the energy fluxes generated by PPM (see Appendix A for more details and for the tests to validate the algorithm). The code was designed to make efficient use of massively parallel computers using the message-passing interface (MPI) for interprocessor communications.

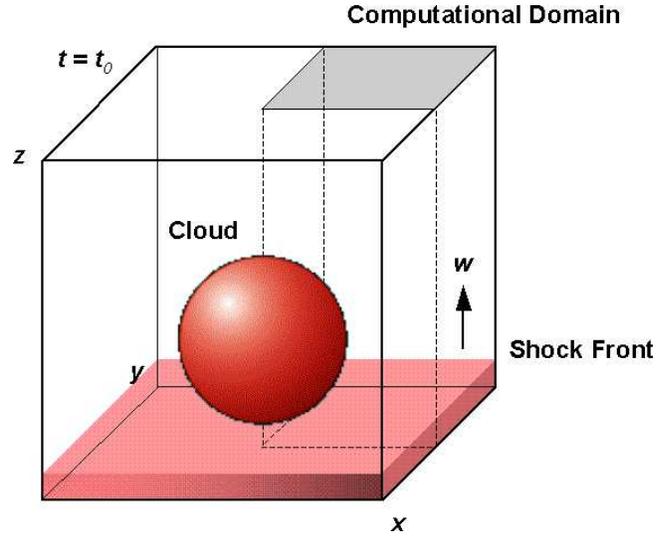

**Figure 1.** Initial geometry of the shock-cloud interaction. The cloud is centered at $(x, y, z) = (0, 0, 0)$. The shock is moving upwards through the ISM with velocity $w$ (see text). Only one quarter of the volume shown is modeled numerically as indicated by a gray patch covering upper right portion of the top face of the domain.

**Table 1.** Summary of the initial physical parameters characterizing the simulations.

|  | Temperature | Density | Velocity |
|---|---|---|---|
| *ISM* | $10^4$ K | 0.1 cm$^{-3}$ | 0.0 |
| *Cloud* | $10^3$ K | 1.0 cm$^{-3}$ | 0.0 |
| *Post-shock medium:* |  |  |  |
| - Mach 30 shock | $1.7 \times 10^6$ K | 0.4 cm$^{-3}$ | 250 km s$^{-1}$ |
| - Mach 50 shock | $4.7 \times 10^6$ K | 0.4 cm$^{-3}$ | 430 km s$^{-1}$ |

### 2.2. Initial and boundary conditions

The initial configuration of our numerical model is schematically shown in Fig. 1. It consists of an unperturbed ambient medium with a spherical cloud in pressure equilibrium with its surrounding; a planar shock moves toward the cloud and starts to interact with it. The unperturbed medium is assumed to be isothermal ($T_{\rm ism} = 10^4$ K) and homogeneous with hydrogen number density $n_{\rm ism} = 0.1$ cm$^{-3}$ (see Tab. 1). The cloud has radius $r_{\rm cl} = 1$ pc and density $n_{\rm cl} = \chi n_{\rm ism}$ ($n_{\rm cl} = 1$ cm$^{-3}$ for density contrast $\chi = 10$); its temperature is determined by pressure balance across the cloud boundary.

The post-shock initial conditions of the ambient medium are given by the strong shock limit (Zel'dovich & Raizer 1966). The post-shock density and velocity are, respectively,

$$n_{\rm psh} = \frac{\gamma + 1}{\gamma - 1}\,n_{\rm ism}, \qquad u_{\rm psh} = \frac{2}{\gamma + 1}w. \tag{6}$$

---

[2] The plasma is allowed to cool down to a nominal threshold of $T \sim 10$ K. This threshold is, however, never reached in the simulations presented here.

[3] The post-shock electron and ion temperatures are considered identical in our model, an hypothesis that is realistic for shocks with the velocities considered here (Rakowski et al. 2003).

Here $w = \mathcal{M} c_{\rm ism}$ is the shock speed, $\mathcal{M}$ is the shock Mach number and $c_{\rm ism}$ is the sound speed in the interstellar medium. The post-shock ion temperature is

$$T_{\rm psh} = \frac{\gamma - 1}{2} \frac{\mu m_H}{2 k_{\rm B}} u_{\rm psh}^2, \tag{7}$$

where $k_{\rm B}$ is the Boltzmann constant. For the $\mathcal{M} = 30$ case (shock speed $w \approx 340$ km s$^{-1}$), $u_{\rm psh} \approx 250$ km s$^{-1}$ and $T_{\rm psh} \approx 1.7 \times 10^6$ K; for the $\mathcal{M} = 50$ case (shock speed $w \approx 570$ km s$^{-1}$), $u_{\rm psh} \approx 430$ km s$^{-1}$ and $T_{\rm psh} \approx 4.7 \times 10^6$ K.

In the 3-D simulations, we solve the hydrodynamic equations in one quadrant of the whole spatial domain (indicated by a dark patch covering the top portion of the volume shown in Fig. 1). The coordinate system is oriented in such a way that the shock front lies in the $(x, y)$ plane and moves in $z$-direction which points through the cloud center. The cloud is initially centered at $(x, y, z) = (0, 0, 0)$ and the computational domain extends $\sim 2$ pc in both the $x$ and $y$ directions, and $\sim 6$ pc in the $z$ direction. At $z = z_{\rm min}$ the variables have been set to the post-shock values while we allowed for free outflow at $z = z_{\rm max}$ and along $x = x_{\rm max}$ and $y = y_{\rm max}$. Both planes cutting through the center of the cloud along $x = x_{\rm min}$ and $y = y_{\rm min}$ have been treated as impenetrable walls. Our assumption of four-fold symmetry in 3-D simulations, aimed at reducing computational cost, leads to a long-wavelength cut-off of the perturbation spectrum experienced by the model object. For example, long-wavelength modes participating in the development of instabilities in the plane parallel to the shock front are not represented in our model. However, we expect that such particular instabilities are not likely to dominate the evolution of the system essentially due to lack of asymmetry (a spherically symmetric cloud, planar shock-wave) and relatively weak waves developing in that plane.

Our 2-D models considered a slab corresponding to the $(x, z)$ plane of the 3-D simulations. The 2-D domain has been extended to $4 \times 12$ pc because in some 2-D simulations the cloud expands faster. We use reflecting boundary conditions at $r = r_{\rm min}$, consistent with the adopted symmetry; a constant inflow is imposed at the lower boundary with free outflow elsewhere.

At the coarsest resolution, the adaptive mesh algorithm used in the FLASH code uniformly covers the 3-D computational domain with a mesh of $2 \times 2 \times 6$ blocks ($4 \times 8$ blocks in the 2-D cases). All the blocks used in the computation have $8^3$ cells ($8^2$ in the 2-D cases). We allow for 5 levels of refinement, with resolution increasing twice at each refinement level. The default refinement criterion adopted follows the changes in density and temperature (Löhner 1987). This grid configuration yields an effective resolution of $\approx 7.6 \times 10^{-3}$ pc at the finest level corresponding to $\approx 132$ zones per cloud radius.

### 2.3. Time-scales

A number of useful dynamical time-scales can be calculated analytically in order to estimate the relative importance of various physical effects. In our discussion we focus on the cloud crushing time, time-scales characteristic of hydrodynamic instabilities, thermal conduction and radiative cooling.

The cloud-crushing time (KMC94), i.e. the characteristic time for the transmitted shock to cross the cloud, is generally defined as

$$\tau_{\rm cc} = \frac{r_{\rm cl}}{w_{\rm cl}} = \frac{\chi^{1/2} r_{\rm cl}}{\beta^{1/2} w} \ . \tag{8}$$

Here $w_{\rm cl} = \beta^{1/2} w / \chi^{1/2}$ is the velocity of the shock transmitted into the cloud (McKee & Cowie 1975, KMC94) and $\beta$ is a parameter of the order of 1 (see Appendix B for a detailed evaluation of $\beta$). For the conditions considered here and $\beta = 1$, the cloud crushing time varies between $\tau_{\rm cc} \approx 9.1 \times 10^3$ yr for $\mathcal{M} = 30$ shock and $\tau_{\rm cc} \approx 5.4 \times 10^3$ yr for $\mathcal{M} = 50$ shock.

The cloud can be subject to both KH and RT instabilities. The KH and RT growth times can be expressed in terms of $\tau_{\rm cc}$ (KMC94):

$$\tau_{\rm KH} \sim \frac{\tau_{\rm cc}}{k_\lambda r_{\rm cl}} \ , \qquad \tau_{\rm RT} \sim \frac{\tau_{\rm cc}}{(k_\lambda r_{\rm cl})^{1/2}} \ . \tag{9}$$

Here $k_\lambda$ is the wave-number of the perturbation. KMC94 showed that the most disruptive wavelengths are those corresponding to $k_\lambda r_{\rm cl} \sim 1$. If not suppressed, these instabilities are responsible for cloud break-up on time-scales comparable with $\tau_{\rm cc}$, eventually leading to efficient mixing of the cloud material with its surrounding and to the final cloud destruction (see also Xu & Stone 1995; Poludnenko et al. 2002).

One of the processes capable of delaying or suppressing destructive action of hydrodynamic instabilities is thermal conduction. Thermal conduction smoothes the temperature and density gradients and, therefore, lowers the efficiency of the KH and RT instabilities. The characteristic time-scale for the conduction is

$$\tau_{\rm cond} = \frac{7}{2(\gamma - 1)} \frac{P}{\kappa(T) T / l^2} \approx 2.6 \times 10^{-9} \frac{n_{\rm H} l^2}{T^{5/2}} \tag{10}$$

where $l$ is a characteristic length of temperature variation. If $\tau_{\rm cond} < \tau_{\rm cc}$, thermal gradients on scales below $l$ are diffused on time-scales shorter than $\tau_{\rm cc}$.

The radiative cooling may also be able to slow down the growth of hydrodynamic instabilities. The cooling leads to the formation of thin dense sheets in the shocked cloud regions (e.g. Falle 1975; Falle 1981). This is the opposite behavior to the diffusive action of the thermal conduction and of the KH and RT instabilities. The cooling time for the shocked gas is

$$\tau_{\rm rad} = \frac{1}{\gamma - 1} \frac{P}{n_{\rm e} n_{\rm H} \Lambda(T)} \approx 2.5 \times 10^3 \frac{T^{3/2}}{n_{\rm e}} \ , \tag{11}$$

where, for temperatures characteristic of our models ($T \approx 10^5 - 10^7$ K), we have approximated the cooling function as $\Lambda(T) \approx 1.6 \times 10^{-19} T^{-1/2}$ erg s$^{-1}$ cm$^3$ (Raymond et al. 1976; Raymond & Smith 1977).

The thermal conduction and the radiative losses are competing effects: the former leads to the heating and evaporation of the cloud, while the latter to the cooling and condensation of the cloud fragments. The radiative losses dominate over the

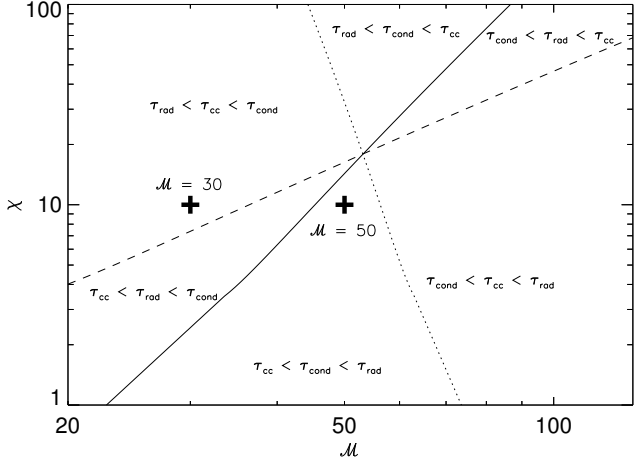

**Figure 2.** $\chi - \mathcal{M}$ parameter space. The lines are derived for length-scale $l = 1$ pc and mark the density contrast for which $[\tau_{\rm cond}]_{\rm psh} = \tau_{\rm cc}$ (dotted line), $[\tau_{\rm rad}]_{\rm cl} = \tau_{\rm cc}$ (dashed line), and $[\tau_{\rm rad}]_{\rm cl} = [\tau_{\rm cond}]_{\rm psh}$ (solid line). The crosses mark the two cases investigated.

effects of the thermal conduction, whenever the cooling time-scale is shorter than the conduction time-scale:

$$\left(\frac{\tau_{\rm rad}}{\tau_{\rm cond}}\right)^{1/2} = \left(\frac{2}{7} \frac{\kappa(T) T}{n_{\rm H}^2 \Lambda(T) l^2}\right)^{1/2} \approx 10^6 \frac{T^2}{n_{\rm H} l} < 1 \ . \quad (12)$$

Note that by setting instead the equality sign into Eq. 12, we derive the Field length scale (Begelman & McKee 1990), i.e. the maximum scale on which the conduction dominates the radiative cooling.

Fig. 2 shows the relationship between the different time-scales[4] as a function of the density contrast $\chi$ and of the shock Mach number $\mathcal{M}$, for a cloud of 1 pc radius. For density contrasts above the dotted line, structures of 1 pc (or below) are subject to the thermal conduction over a time-scale shorter than $\tau_{\rm cc}$; for density contrasts above the dashed line, the shocked cloud material is subject to the radiative losses over a time-scale shorter than $\tau_{\rm cc}$; the solid line divides the (left) region dominated by the radiative cooling from the (right) region dominated by the thermal conduction. In Appendix B, we derive the cooling time-scale behind the shock transmitted into the cloud, $[\tau_{\rm rad}]_{\rm cl}$, and the thermal conduction time-scale in the shocked ambient medium, $[\tau_{\rm cond}]_{\rm psh}$, expressed in terms of $\mathcal{M}$ and $\chi$.

Fig. 2 shows that, in a $\mathcal{M} = 30$ case, $[\tau_{\rm rad}]_{\rm cl} < \tau_{\rm cc} < [\tau_{\rm cond}]_{\rm psh}$ and, therefore, we expect that the shock transmitted into the cloud is strongly radiative and that its evolution is dominated by the energy losses on time-scales shorter than the cloud crushing time. On the other hand, the thermal conduction dominates over the radiative losses in a $\mathcal{M} = 50$ case; the cloud is expected to evaporate on a time-scale comparable with $\tau_{\rm cc}$ (since this case is located close to the dotted line; $[\tau_{\rm cond}]_{\rm psh} \gtrsim \tau_{\rm cc}$).

Ferrara & Shchekinov (1993) pointed out that the conductive fronts do not induce relevant dynamical effects if $\tau_{\rm cond}$ is much shorter than the dynamical response time of the system, which can be approximated as the sound crossing time of the cloud: $\tau_{\rm dyn} \sim r_{\rm cl}/c_{\rm s}$. From Eq. 10, it is easy to show that

$$\frac{\tau_{\rm cond}}{\tau_{\rm dyn}} \approx 3 \times 10^{-5} \frac{n_{\rm H} l^2}{r_{\rm cl} T^2} \ . \quad (13)$$

Considering the hydrogen number density of the shocked cloud $n_{\rm H} = 4 \ {\rm cm}^{-3}$, $r_{\rm cl} = 1$ pc, and a length scale $l = 1$ pc, the conductive time-scale is shorter than the dynamical time-scale for $T > 2 \times 10^7$ K. In the cases considered here, however, the temperatures are lower and the evolution of conductive fronts is expected to influence gas dynamics. In the saturated regime, the thermal conduction time scale becomes even longer, strengthening the above argument.

The saturated heat flux (Eq. 4) provides a lower limit on the conduction time-scale:

$$\tau_{\rm sat} = \frac{l}{5\phi(\gamma - 1)} \left(\frac{2k}{\mu m_{\rm H}}\right)^{-1/2} T^{-1/2} \quad (14)$$

and an upper limit on the propagation velocity of the conduction front:

$$v_{\rm sat} = \frac{l}{\tau_{\rm sat}} = 5\phi(\gamma - 1) \left(\frac{2k}{\mu m_{\rm H}}\right)^{1/2} T^{1/2} \ . \quad (15)$$

In the case of strong shocks subject to heat transfer, a thermal precursor develops if the propagation velocity of the conduction front is larger than the shock speed $w$ (Zel'dovich & Raizer 1966). By combining Eqs. 6, 7, and 15, the condition for $v_{\rm sat} < w$ reads

$$\phi < \frac{1}{5\sqrt{2}} \frac{(\gamma + 1)}{(\gamma - 1)^{3/2}} \approx 0.7 \ . \quad (16)$$

Since we use $\phi = 0.3$, typical of a fully ionized cosmic gas, in all the simulations (see Sect. 2.1), $v_{\rm sat} < w$ and no thermal precursor develops. Our choice turns out to be consistent with the fact that no precursor is observed in young and middle aged SNRs.

## 3. Results

### 3.1. The simulations

We consider two examples of shock-cloud interaction with different Mach number ($\mathcal{M} = 30$ and $\mathcal{M} = 50$), to analyze the dynamics when either the heat conduction or the radiation plays a dominant role. The effects of the thermal conduction and of the radiation in the shock-cloud interaction are also investigated by comparing these models with other models without conduction and radiation.

The models neglecting both thermal conduction and radiation have been computed in a 3-D Cartesian coordinate system $(x, y, z)$, in order to describe well the hydrodynamic instabilities at the boundaries of the shocked cloud (e.g. Xu & Stone 1995). On the other hand, the heat conduction is known to damp rapidly the hydrodynamic instabilities; therefore, a 2-D cylindrical coordinate system $(r, z)$ is enough to describe the shock-cloud interaction including the thermal conduction: indeed we anticipate that, in the latter case, there are no complex instabilities to follow.

---
[4] Ferrara & Shchekinov (1993) derived an analogous diagram in the context of the dynamics of conductive/cooling fronts.

**Table 2.** Summary of the 3- and 2-D shock-cloud simulations.

| Run | Geometry | Res.[a] | $\mathcal{M}$ | $\chi$ | therm. cond. & rad. losses |
|---|---|---|---|---|---|
| HY1 | 3-D cart. $(x, y, z)$ | 105 | 50 | 10 | no |
| HY2 | 3-D cart. $(x, y, z)$ | 132 | 50 | 10 | no |
| HTR30 | 2-D cyl. $(r, z)$ | 132 | 30 | 10 | yes |
| HTR50 | 2-D cyl. $(r, z)$ | 132 | 50 | 10 | yes |

[a] In number of zones per cloud radius, $r_{\rm cl}$.

KMC94 showed that the purely hydrodynamic shock-cloud problem (without thermal conduction and radiation) is independent of the Mach number of the shock (the so-called Mach-scaling) and invariant under the scaling

$$t \to t\mathcal{M}, \quad u \to u/\mathcal{M}, \quad T \to T/\mathcal{M}^2, \qquad (17)$$

where $t$ is the time, $\mathcal{M}$ the Mach number, $u$ the gas velocity, and $T$ the temperature, with distance, density, and pre-shock pressure left unchanged, and provided that $\mathcal{M} \gg 1$. Therefore, only one simulation without thermal conduction and radiation is required for our purposes and the results are representative of all the cases examined, provided that $\mathcal{M} \gg 1$.

We run two 3-D simulations with different spatial resolution to check if the adopted resolution is sufficient to capture the basic cloud evolution over the time interval considered. Both simulations have a spatial resolution higher than previously obtained in 3-D in the shock-cloud problems: they have more than 100 zones per cloud radius, following the prediction of KMC94 and Mac Low et al. (1994) that this is the minimum spatial resolution for adequate description of all physical quantities. A summary of all the simulations in this paper is given in Table 2.

We first briefly describe the relevant aspects of the shock-cloud interaction without thermal conduction and radiative losses. We then investigate the effects of both physical processes on the dynamics by comparing the above models with other models that take full account of conduction and radiation.

### 3.2. Evolution neglecting thermal conduction and radiation

We simulate the impact of the Mach 50 shock on the cloud: as discussed above, the case of the $\mathcal{M} = 30$ shock can be derived, through simple scaling, from the $\mathcal{M} = 50$ case and, therefore, no specific simulation is required.

The evolution of the $\mathcal{M} = 50$ shock-cloud interaction is illustrated for the highest resolution simulation (run HY2 with a resolution of $\sim 132$ zones per cloud radius) in the 3-D visualizations of Fig. 3. As in previous 2-D and 3-D studies (KMC94, Xu & Stone 1995, and references therein), we divide the early shock-cloud interaction into four stages:

1. *Initial phase* [$t < 0.64\,\tau_{\rm cc}$].
   The shock encounters the cloud and their interaction leads to the formation of transmitted (into the cloud) and reflected (into the shocked ISM) shocks (panel A in Fig. 3). The darkest portion of the cloud in Fig. 3 is the shocked cloud material. In our simulations, the temperature of the shock transmitted into the cloud is $\sim 10^6$ K. The reflected (bow) shock propagates into the shocked ISM with a temperature of $\sim 6 \times 10^6$ K. During this phase, the cloud is compressed by the transmitted shock. This phase lasts approximately the time taken by the shock in the ambient medium to sweep across the cloud $\sim 2r_{\rm cl}/w = 2/\chi^{1/2}\,\tau_{\rm cc} \simeq 0.64\,\tau_{\rm cc}$ (KMC94, Xu & Stone 1995).

2. *Shock compression* [$0.64\,\tau_{\rm cc} < t < 1.1\,\tau_{\rm cc}$].
   The flow around the cloud converges on the symmetry axis ($z$-axis) and the ambient post-shock pressure compresses the cloud from all directions. By the end of this stage ($t \simeq \tau_{\rm cc}$, see panel B in Fig. 3), the primary SNR shock undergoes a conical self-reflection (Tenorio-Tagle & Różyczka 1984) and a reverse shock is driven back into the cloud (Woodward 1976).

3. *Re-expansion phase* [$1.1\,\tau_{\rm cc} < t < 1.9\,\tau_{\rm cc}$].
   The combined action of the high pressure inside the cloud, due to the transmitted shock, and of the rarefaction of the ambient gas, due to the reflection of the SNR shock, leads to the expansion of the cloud (Woodward 1976, KMC94, Xu & Stone 1995 and references therein). At the same time, the SNR shock deposits vorticity at the cloud surface, and triggers the development of hydrodynamic instabilities (Saffman & Baker 1979, KMC94; see panel C in Fig. 3).

4. *Cloud destruction* [$t > 1.9\,\tau_{\rm cc}$].
   This phase is dominated by hydrodynamic instabilities (see panel D in Fig. 3). The complex velocity field leads to a complex pattern of filaments and cloud fragments in a region of highly non-uniform density. Ultimately, the action of the hydrodynamic instabilities destroys the cloud after several $\tau_{\rm cc}$ (KMC94).

### 3.3. The role of thermal conduction and radiation

#### 3.3.1. The Mach 50 shock case (HTR50)

Figs. 4 and 5 show the evolution of the mass density and of the temperature, respectively, in a 2-D section of the $(x, z)$ plane in the simulations HY2 (left half panels) and HTR50 (right half panels).

During the first two stages ($t < 1.1\,\tau_{\rm cc}$), the whole front face of the cloud, overrun by the shock and prone to hydrodynamic instabilities, is strongly diffused and the shocked cloud material is quickly heated. The cloud stripping, present in HY2, is masked by the evaporation process as well. A transition region from the inner part of the cloud to the ambient medium gradually grows during the evolution, after the expansion of the cloud. In such a region, the density and temperature gradients vary very smoothly in the radial direction. As we will discuss below in more detail, the reflected shock in HTR50 is slightly stronger and cooler than in HY2 as some fraction of its thermal energy is conducted into the evolving cloud boundary and some fraction of the cloud material is mixed in the surrounding medium. During the third stage ($1.1\,\tau_{\rm cc} < t < 1.9\,\tau_{\rm cc}$), the cloud is progressively heated up to the temperature of the surrounding medium. After $\sim 2.2\ \tau_{cc}$ the cloud has density contrast $\chi \sim 1$ with respect to the surrounding medium. Note that, although the density and the temperature of the cloud are indistinguishable from those of the surrounding medium at the end of the simulation, the cloud material is not fractioned in small cloudlets as in HY2 (see the contour enclosing the cloud material).

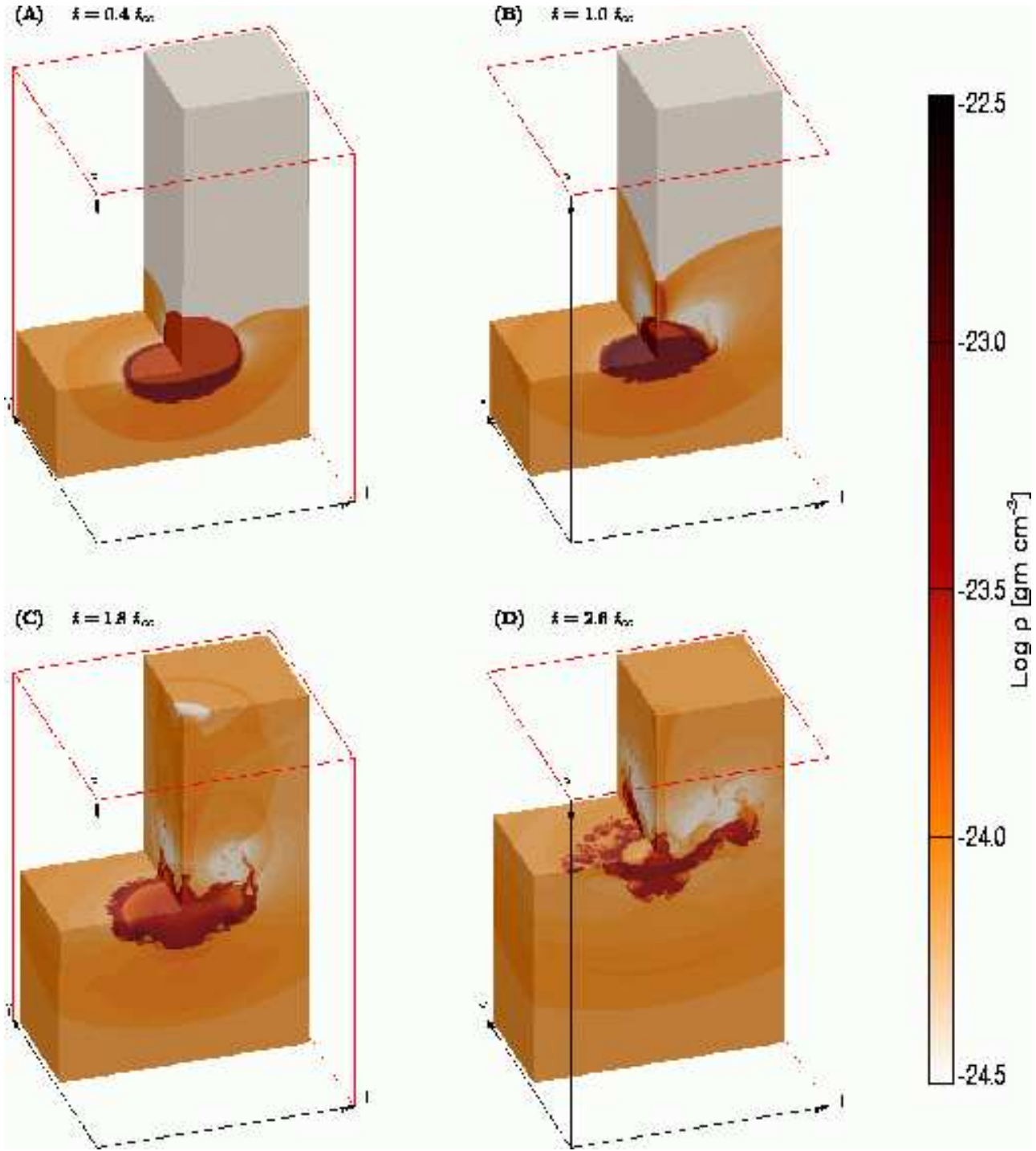

**Figure 3.** 3-D visualizations of the mass density evolution during the shock-cloud interaction at selected times in units of $\tau_{cc}$. The gray scale shows the density distributions, in log scale. The box is $4 \times 4 \times 6$ pc$^3$. The calculation is the one neglecting the thermal conduction and the radiative losses and at the highest spatial resolution (run HY2, see Table 2).

Figs. 4 and 5 also show that the thermal conduction influences the propagation speed of the shocks generated during the evolution: the shock transmitted into the cloud is faster and the reflected bow shock is slower than those generated in HY2. In addition, we note that the self-reflected part of the primary shock is slower in HTR50 than in HY2. These results can be understood looking at Fig. 6, which show the mass density and temperature profiles along the symmetry axis, $z$, in HY2

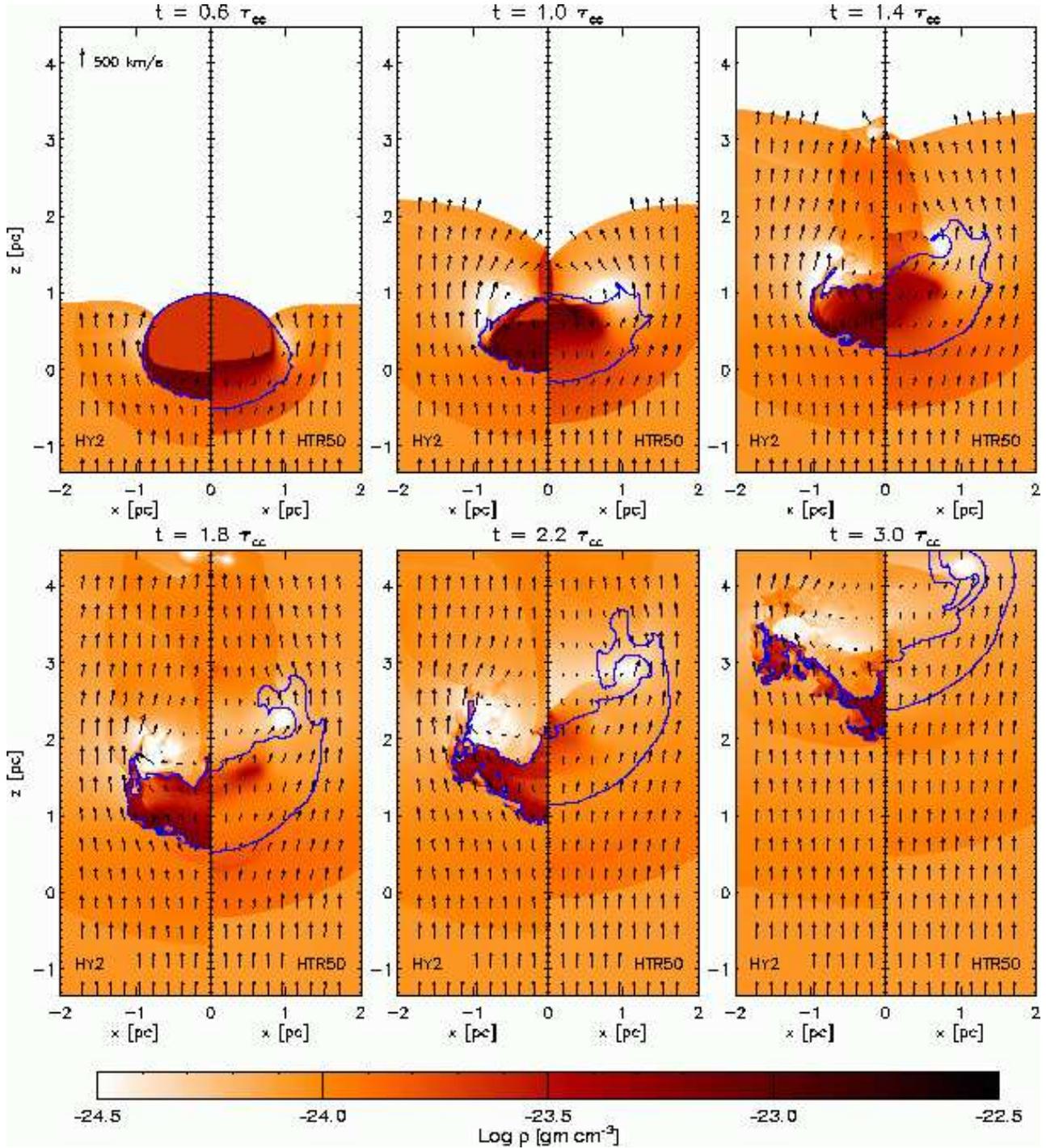

**Figure 4.** 2-D sections in the $(x, z)$ plane of the mass density distribution (gm cm$^{-3}$), in log scale, in the simulations HY2 (left half panels) and HTR50 (right half panels), sampled at the labeled times in units of $\tau_{cc}$. The velocity arrows scale linearly with respect to the reference velocity shown in the upper left panel and corresponding to 500 km s$^{-1}$. The contour encloses the cloud material.

and HTR50. In HTR50 we see the progressive heating of the shocked cloud material and its evaporation in the surrounding medium (see panels at $t = 0.6\,\tau_{cc}$ and $1\,\tau_{cc}$), driven by the thermal conduction. As a result, the material behind the transmitted shock in HTR50 is at higher temperature and at lower density than that in HY2. Since the propagation velocity of a shock depends on the temperature of the post-shock plasma (see Eqs. 6 and 7; Zel'dovich & Raizer 1966), the transmitted shock in HTR50 is faster than that in HY2. On the other hand, the material behind the reflected shock in HTR50 is cooler than that in HY2 because a fraction of its thermal energy has been conducted into the cloud, and denser than that in HY2 because a fraction of the cloud material has evaporated into the surrounding shocked medium (see Fig. 6). As a consequence, the reflected shock in HTR50 is slower than that in HY2. Analogous

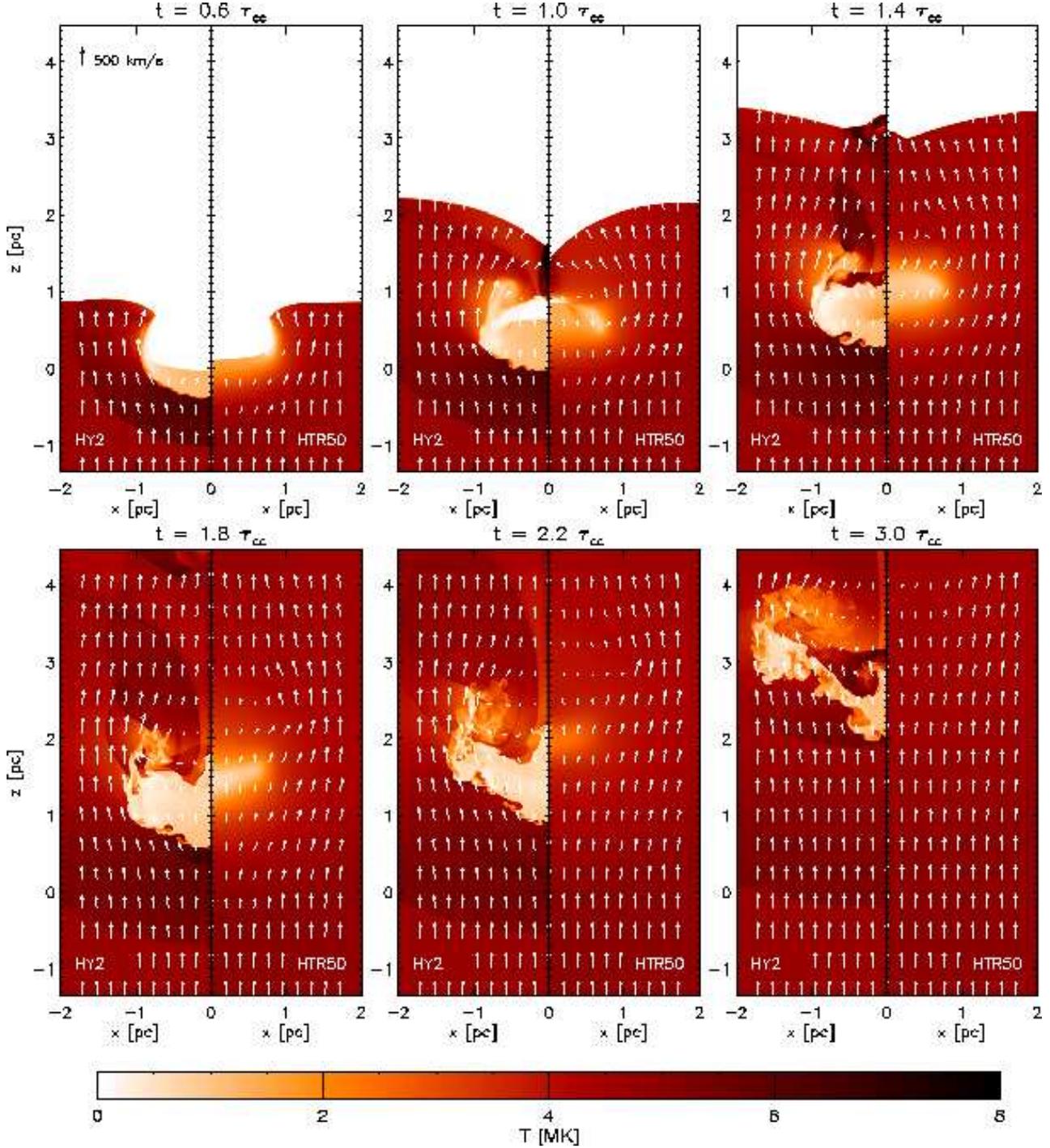

**Figure 5.** As in Fig. 4 for the plasma temperature distribution (MK).

considerations explain why the self-interacting primary shock in HTR50 is slower than that in HY2.

### 3.3.2. The Mach 30 shock case (run HTR30)

We compare the evolution of the shock-cloud interaction derived in simulation HTR30 with that derived in HY2, after scaling the velocity and temperature distributions derived in the latter, according to the transformations shown in Eq. 17. Figs. 7 and 8 compare the evolution of the density and temperature distribution, respectively, in these two runs.

Just as in the simulation with $\mathcal{M} = 50$, during the first two stages ($t < 1.1\,\tau_{\rm cc}$), the thermal conduction limits the development of the dynamical instabilities. On the other hand, for $\mathcal{M} = 30$, the shocked cloud evolves in a dense and cold structure: the strong cooling in the post-shock cloud region results in the rapid accumulation of the cooled material in a thin dense shell (Falle 1981) as well as in the substantial weakening of the transmitted shock. The shell forms very quickly (at $t \sim 0.6\,\tau_{cc}$ it is already there) in agreement with the cooling time $\tau_{rad} \sim 0.4\,\tau_{cc}$ (see Appendix B). On the other hand, a diluted outer part of the cloud starts to develop a hot corona surrounding the dense shell and characterized by particle density $n \approx 0.4$ cm$^{-3}$ and $T \approx 8 \times 10^5$ K. From Eq. 12 and from the above values of the density and the temperature, we derive that the evolution of this corona is dominated by the thermal conduction. At $t \sim \tau_{cc}$, the shock propagating in the ISM has enveloped the cloud, focused behind it, and started to enter from the rear: it forms, therefore, its own dense shell and a transient ring-like shell configuration forms in the time interval $\tau_{cc} < t < 1.2\,\tau_{cc}$. During the following phases ($t > 1.1\,\tau_{cc}$), the cooling-dominated cloud material fragments into dense, cold, and compact filaments which survive until the end of our simulation, whereas the hot corona gradually evaporates under the effect of the thermal conduction. The strong cooling leads to cool and dense material accumulating along the symmetry axis in Figs. 7 and 8. Note that the details of the plasma cooling to such low temperatures depend, however, on the cooling function adopted in our computations which is appropriate to temperatures $T > 4 \times 10^3$ K, and on the numerical resolution which affect the peak density and hence the cooling efficiency of the gas.

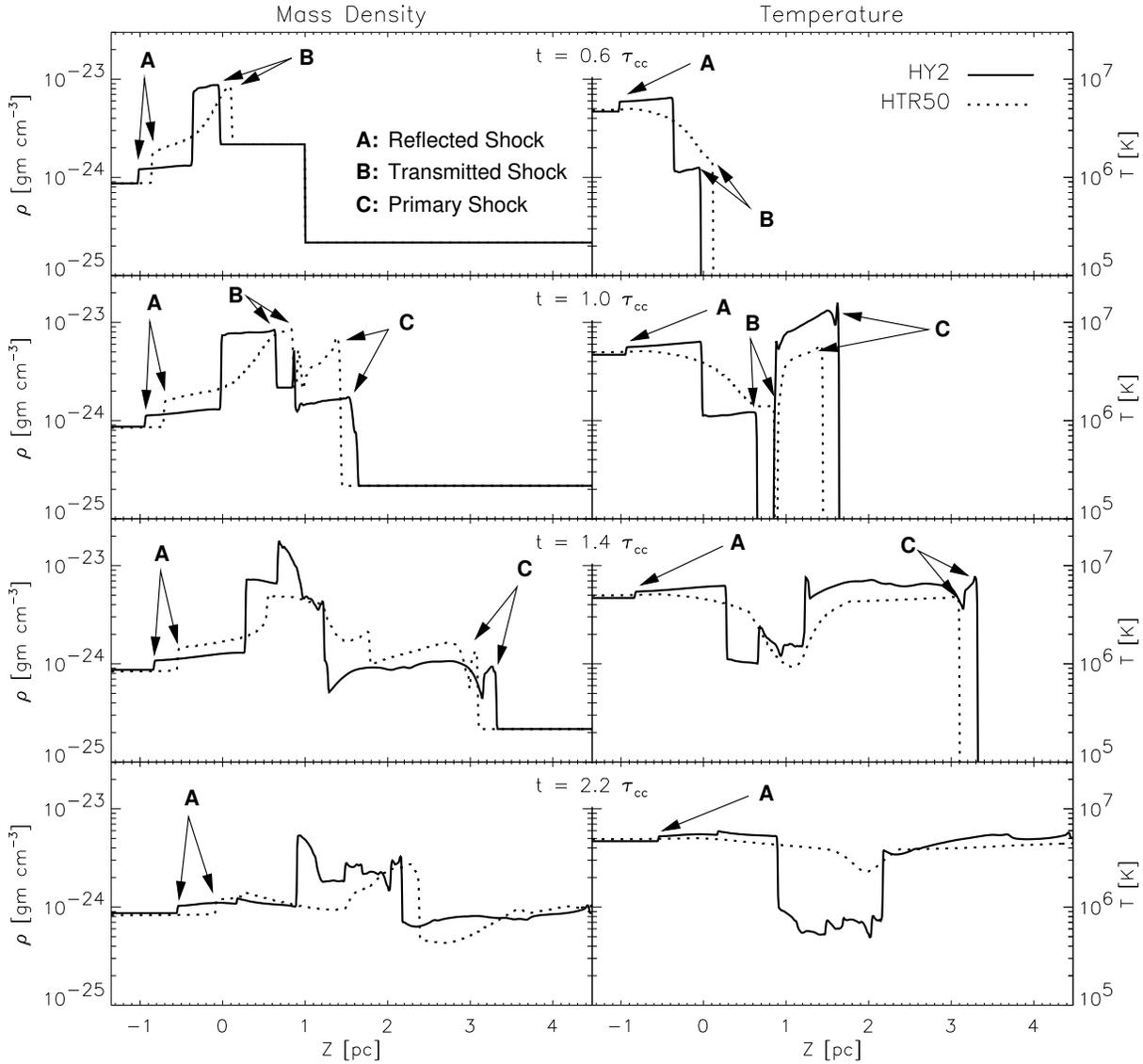

**Figure 6.** Mass density (left panels) and temperature (right panels) profiles along the symmetry axis, $z$, in the runs HY2 (solid lines) and HTR50 (dotted lines) at selected times in units of $\tau_{cc}$. The arrows mark the reflected (A), transmitted (B) and primary (C) shocks for the two simulations.

The tracer defined in Eq. 5 and the Eq. 12 allow us to investigate further the efficiency of the radiative losses. We first identify zones consisting of the original cloud material by more than 90%. We then quantify the mass fraction of this material ($M/M_{cl0}$ where $M_{cl0}$ is the cloud mass at the beginning of the interaction) dominated by the radiative losses and the mass fraction dominated by the thermal conduction as a function of time, applying Eq. 12 in each of those zones (see Fig. 9). We find that the thermal collapse starts at $t \sim 0.4\,\tau_{cc}$, in agreement with our previous results (see also Appendix B). The cooling process is very efficient throughout the cloud and $\sim 80\%$ of the initial cloud material is subject to radiative cooling at $t > \tau_{\rm cc}$. On the other hand, as mentioned above, not all of the cloud is dominated by radiative losses: $\sim 10\%$ of the initial cloud mass forms the hot thermally conducting corona located around the cooling-dominated region. The remaining 10% of the initial cloud mass is mixed together with the ambient medium. During the shock-cloud interaction, the hot corona expands and gradually evaporates under the effect of thermal conduction, whereas the cold core collapses and fragments in cloudlets under the effect of radiation.

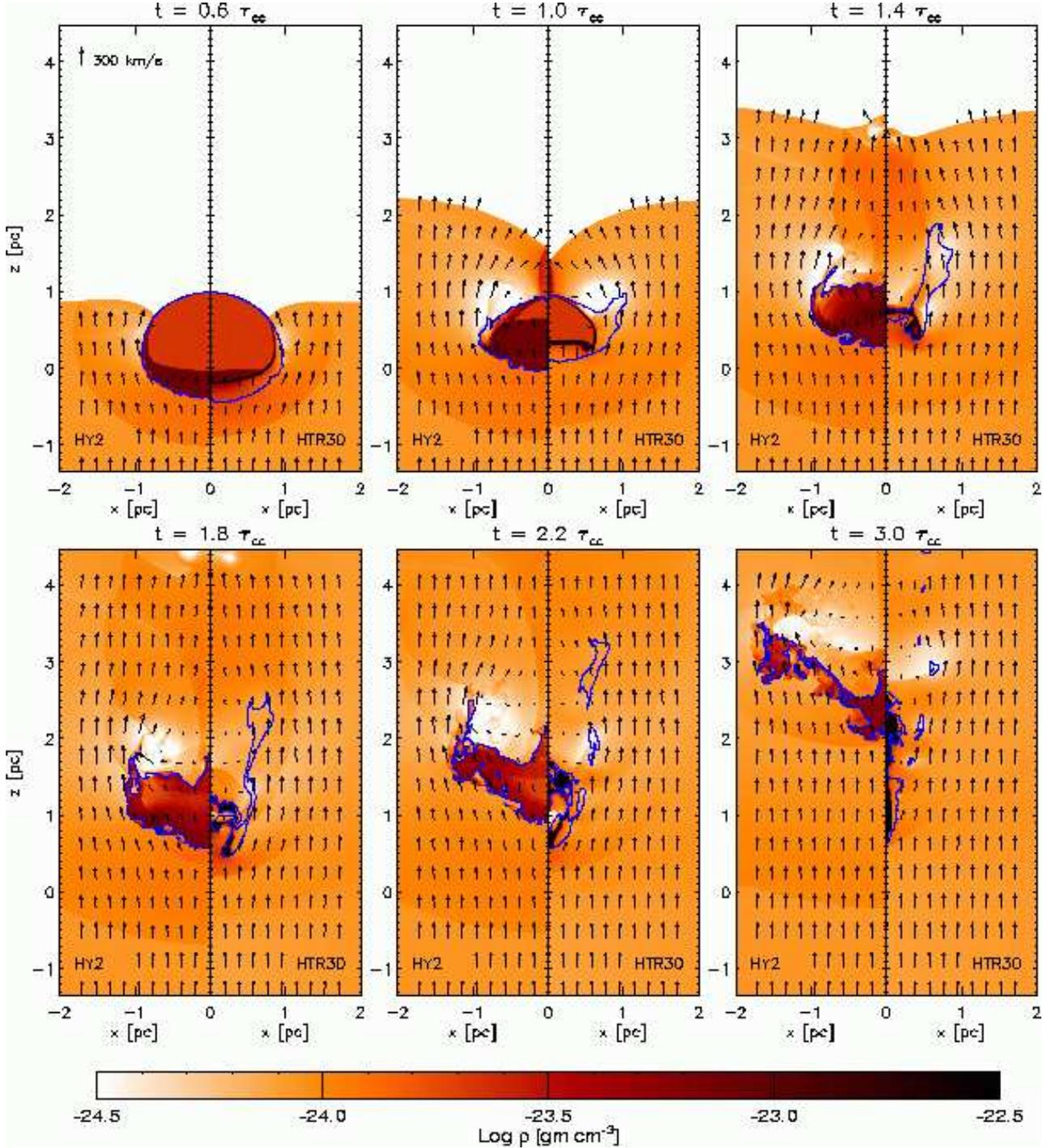

**Figure 7.** As Fig. 4 for the Mach 30 case. Note that the velocity field calculated in the run HY2 has been scaled by the factor 3/5, according to the Mach scaling (see Eq. 17). The velocity arrows scale linearly with respect to the reference velocity shown in the upper left panel and corresponding to 300 km s$^{-1}$. The contour encloses the cloud material.

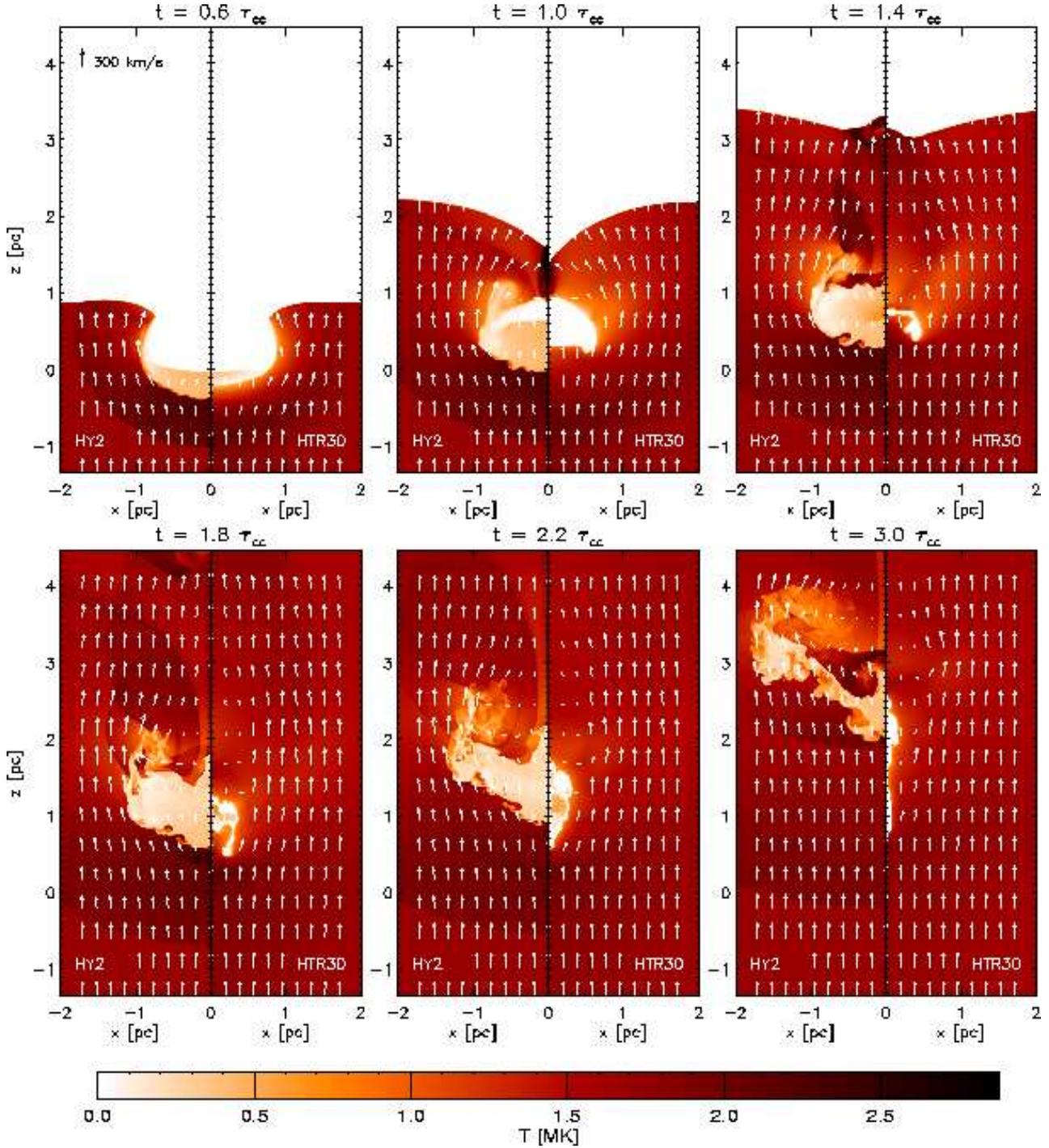

**Figure 8.** As Fig. 5 for the Mach 30 case. The velocity field and the temperature distribution calculated in the run HY2 have been scaled by the factor 3/5 and $(3/5)^2$, respectively, according to the Mach scaling (see Eq. 17).

### 3.4. Global quantities

In order to study quantitatively the global evolution of the gas cloud, we consider several diagnostic variables. Eq. (5) allows us to trace the cloud material in the ambient gas. We use such a tracer to identify zones whose content is the original cloud material by more than 90%; then we define the cloud mass, $M_{\rm cl}$, as the total mass in these zones, and the cloud volume, $V_{\rm cl}$, as the total volume occupied by these zones, namely:

$$M_{\rm cl} = \int_{V(C_{\rm cl}>0.9)} C_{\rm cl}\rho \, dv, \quad V_{\rm cl} = \int_{V(C_{\rm cl}>0.9)} dv \quad (18)$$

where the integral is done on zones with $C_{\rm cl} > 0.9$.

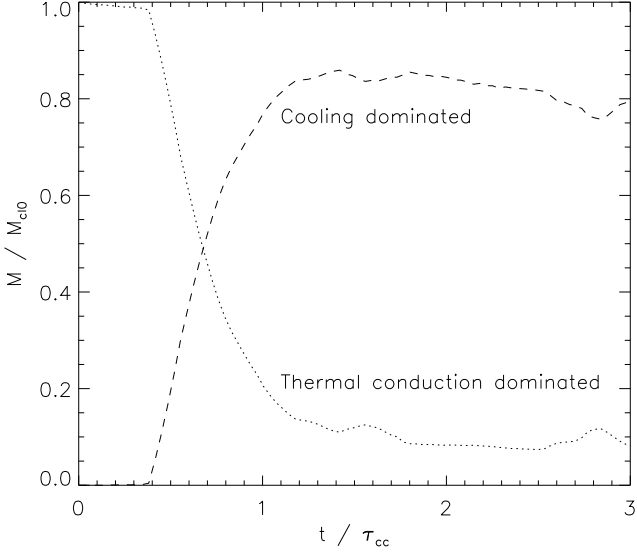

**Figure 9.** Mass fraction of the initial cloud material dominated by the radiative losses (dashed line) or by the thermal conduction (dotted line) in run HTR30.

Following Xu & Stone (1995), we study the mixing of the cloud material with the ambient gas, defining the mixing fraction, $f_{\rm mix}$, as

$$f_{\rm mix} = \frac{M_{\rm mix}}{M_{\rm cl0}} = \frac{1}{M_{\rm cl0}} \int_{V(0.1<C_{\rm cl}<0.9)} C_{\rm cl}\rho \, dv \qquad (19)$$

where $M_{\rm mix}$ is the cloud mass in zones which contain more than 10% and less than 90% of the cloud material (the integral is over zones with $0.1 < C_{\rm cl} < 0.9$) and $M_{\rm cl0}$ is the cloud mass at the beginning of the interaction.

The cloud mass and volume in Eq.(18) allow us to derive an average particle density of the cloud as

$$\langle n_{\rm e} \rangle_{\rm cl} = \frac{M_{\rm cl}}{\mu m_H V_{\rm cl}} \; . \qquad (20)$$

We also define an average mass-weighted temperature and an average mass-weighted velocity of the cloud in the direction of shock propagation, respectively, as

$$\langle T \rangle_{\rm cl} = \frac{\int_{V(C_{\rm cl}>0.9)} C_{\rm cl}\rho T \, dv}{\int_{V(C_{\rm cl}>0.9)} C_{\rm cl}\rho \, dv} \qquad (21)$$

$$\langle u \rangle_{\rm cl} = \frac{\int_{V(C_{\rm cl}>0.9)} C_{\rm cl}\rho u_z \, dv}{\int_{V(C_{\rm cl}>0.9)} C_{\rm cl}\rho \, dv} \qquad (22)$$

where, again, we integrate on zones with $C_{\rm cl} > 0.9$ and $u_z$ is the velocity component in the $z$-direction.

Fig. 10 plots the evolution of the global quantities for all the simulations of Table 2: the normalized cloud mass $M_{\rm cl}/M_{\rm cl0}$, the normalized cloud volume, $V_{\rm cl}/V_{\rm cl0}$ (where $V_{\rm cl0}$ is the initial cloud volume), the mixing fraction, $f_{\rm mix}$, the average cloud density normalized to the density of the shocked ambient gas, $\langle n_{\rm e}\rangle_{\rm cl}/n_{\rm psh}$, the average mass-weighted cloud temperature normalized to the temperature of the shocked ambient gas, $\langle T \rangle_{\rm cl}/T_{\rm psh}$, and the average mass-weighted $z$-component of the cloud velocity relative to the shocked ambient gas, $\langle u \rangle_{\rm cl}/u_{\rm psh}$.

The dashed and the solid lines in Fig. 10 mark the results for the simulations without both thermal conduction and radiation (runs HY1 and HY2). The discrepancies between the 3-D low- and high-resolution results are very small, indicating that both our 3-D simulations have enough resolution to capture the dominant dynamical effects present in the evolution, as predicted by KMC94 and Mac Low et al. (1994). Fig. 10 also shows that, when the thermal conduction and the radiation are negligible, the mass loss rate of the cloud becomes significant after one $\tau_{\rm cc}$ (i.e. after the hydrodynamic instabilities have fully developed at the cloud boundary) and $\sim 50\%$ of the cloud mass is contained in mixed zones at $t = 3\,\tau_{\rm cc}$. During the first two stages ($t < 1.1\,\tau_{\rm cc}$), the cloud is progressively compressed: its volume decreases down to 30% of the initial value, its average density and temperature increase up to 3 cm$^{-3}$ ($\sim 8\,n_{\rm psh}$) and $8 \times 10^5$ K ($\sim 0.18\,T_{\rm psh}$), respectively. During the third phase ($1.1\,\tau_{\rm cc} < t < 1.9\,\tau_{\rm cc}$), $V_{\rm cl}$ re-expands back to 50% of the initial volume, leading to a decrease of both $\langle n_{\rm e}\rangle_{\rm cl}$ and $\langle T \rangle_{\rm cl}$ to $\sim 2$ cm$^{-3}$ ($\sim 5\,n_{\rm psh}$) and $\sim 6 \times 10^5$ K ($0.12\,T_{\rm psh}$), respectively. In the last phase, the cloud is first slightly compressed (at $1.9\,\tau_{\rm cc} < t < 2.2\,\tau_{\rm cc}$) by the interaction with the "Mach disk" formed during the reflection of the primary shock at the symmetry axis, and both $\langle n_{\rm e}\rangle_{\rm cl}$ and $\langle T \rangle_{\rm cl}$ increase; then the average cloud density slightly decreases, while $\langle T \rangle_{\rm cl}$ continues to increase and $V_{\rm cl}$ to decrease, because of the mixing of the cloud material with the ambient medium. During the whole evolution, the cloud is continuously accelerated: the average cloud velocity increases up to $\sim 0.7\,u_{\rm psh}$ at $t \sim 3\,\tau_{\rm cc}$.

The dotted lines of Fig. 10 mark the results for HTR50. The mass loss rate of the cloud in this case is almost constant during the whole simulation ($t < 3\,\tau_{\rm cc}$), while in HY2 (and HY1) it becomes significant only after $\sim \tau_{\rm cc}$. This different behavior is due to the different mechanism of cloud mass loss: in HTR50 the mass loss comes from the cloud evaporation driven by the thermal conduction, whereas in HY2 the hydrodynamic instabilities ablate the cloud after $\sim \tau_{\rm cc}$ (see Sect. 2.3). During the first two stages ($t < 1.1\,\tau_{\rm cc}$), therefore, the mass loss rate in HTR50 is more efficient than that in HY2 and, at $t = 1.1\,\tau_{\rm cc}$, $\sim 5\%$ of the cloud mass has been already mixed with the ambient medium ($\sim 0\%$ in run HY2). On the other hand, during the third and fourth stages, in HTR50 the mass loss rate is less efficient than that in HY2 and, at $t = 3\,\tau_{\rm cc}$, only $\sim 10\%$ of the cloud mass is contained in mixed zones ($\sim 50\%$ in run HY2).

In HTR50, during the first stage ($t < 0.65\,\tau_{\rm cc}$), the cloud is compressed, the volume slightly decreases down to 80% of the initial value and the average density of the cloud slightly increases up to $\sim 1.2$ cm$^{-3}$ ($\sim 3\,n_{\rm psh}$). During this phase the cloud is heated efficiently by the thermal conduction and its average temperature increases rapidly to $T \sim 1.6 \times 10^6$ K ($\sim 0.35\,T_{\rm psh}$). As a consequence, the pressure inside the cloud

14 S. Orlando et al.: Crushing of Interstellar Gas Clouds in SNRs. I.

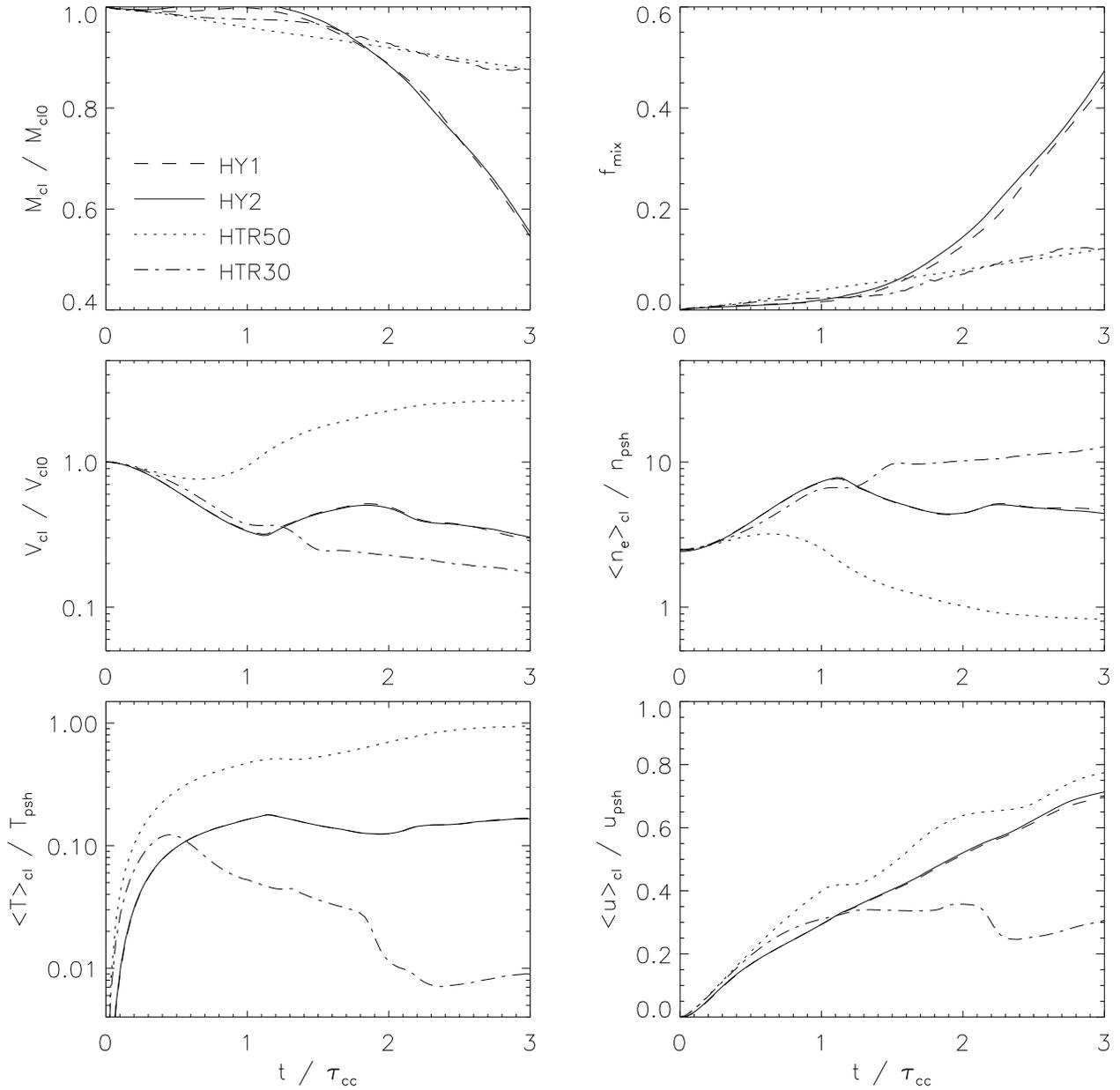

**Figure 10.** Evolution of the global properties of the gas cloud defined in Sect. 3.4 for runs HY1 and HY2 which neglect the thermal conduction and the radiation (dashed and solid lines), and for runs HTR50 and HTR30, which instead include the thermal conduction and the radiation (dotted and dash-dotted lines).

increases and, at $t \sim 0.64\,\tau_{cc}$, the cloud expands again, earlier than in HY2 ($t \sim 1.1\,\tau_{cc}$) until $t < 2.2\,\tau_{cc}$: $V_{cl}$ increases by a factor 2.5 of the initial volume, $\langle n_e \rangle_{cl}$ gradually decreases to $0.3 - 0.4$ cm$^{-3}$, namely the mass density of the surrounding ambient medium ($\langle n_e \rangle_{cl} \approx n_{psh}$). During this phase, the average cloud temperature, $\langle T \rangle_{cl}$, keeps increasing up to $3.8 \times 10^6$ K ($\sim 0.8\,T_{psh}$).

In the last phase ($t > 2.2\,\tau_{cc}$), the cloud volume is almost constant; $\langle n_e \rangle_{cl}$ slightly decreases down to $\sim 0.32$ cm$^{-3}$ ($\sim 0.8\,n_{psh}$), while $\langle T \rangle_{cl}$ keeps increasing up to $\sim 4.5 \times 10^6$ K ($\sim 0.95\,T_{psh}$), because of the thermal conduction. During the whole evolution, the cloud is continuously accelerated up to $\sim 0.8\,u_{psh}$, i.e. more than in HY2, because the cloud has a larger volume and offers, therefore, a larger surface to the pressure of the shock front.

The dot-dashed lines mark the results for $\mathcal{M} = 30$ (HTR30). In this case, the trend of the cloud mass loss rate is similar to the one in HTR50, indicating that the mass loss is again driven by the thermal conduction rather than by hydrodynamic instabilities. In fact, although the radiative losses are dominant in HTR30, a small fraction of the cloud forms a hot corona (in which the thermal conduction is the dominant process) around the cooling-dominated portion of the cloud (see Sect. 3.3.2). Since the mass exchange between the cloud and the ambient medium occurs at the cloud boundary coincident with the boundary of the corona, the mass loss rate is again regulated by the thermal conduction.

During the first stage ($t < 0.5\,\tau_{\rm cc}$), the evolution of the cloud is similar to that of the other simulations: the cloud is compressed, rapidly heated up to $2 \times 10^5$ K ($\sim 0.1\,T_{\rm psh}$) due both to the shock compression and to the thermal conduction, and accelerated up to $0.2\,u_{\rm psh}$. At $t \approx 0.5\,\tau_{\rm cc}$, the transmitted shock becomes strongly radiative (see Sect. 2.3). At variance with the other simulations, there is no re-expansion phase in run HTR30 and the cloud volume decreases down to 20% of the initial value by the end of the simulation. The average density of the cloud has increased by a factor 5, at $t \sim 3\,\tau_{\rm cc}$. In spite of the cloud compression, during this phase the average cloud temperature decreases due to the efficient radiative cooling and, at $t \sim 3\,\tau_{\rm cc}$, $\langle T \rangle_{\rm cl} \approx 1.4 \times 10^4$ K ($\sim 0.008\,T_{\rm psh}$). The cloud is first accelerated up to $0.3\,u_{\rm psh}$ at $t \sim \tau_{\rm cc}$, and thereafter $\langle u \rangle_{\rm cl}$ ranges between 0.2 and $0.4\,u_{\rm psh}$, because the cloud collapses and offers therefore a smaller surface to the pressure of the shock front.

Note that the emergence of a dense and cold interstellar gas phase (as in the $\mathcal{M} = 30$ case) or the evaporation of the whole cloud (as in the $\mathcal{M} = 50$ case), besides the shock Mach number, also depends on the density and dimensions of the cloud. For instance, we expect that a higher initial cloud density would lead to stronger radiative losses and, therefore, to enhance their role in the evolution of the shock-cloud interaction: cooling-dominated regions may exist even at high temperatures, in appropriate physical conditions. As shown in Fig. 2, the evolution of the shocked cloud may be dominated by the radiative losses even in the $\mathcal{M} = 50$ shock case if the cloud has a density contrast $\chi = 30$ (corresponding to a particle density $n_{\rm cl} \sim 3$ cm$^{-3}$). As for the cloud dimensions, we expect that, for moderate cloud densities ($n_{\rm cl} \sim 1$ cm$^{-3}$) and moderate shock Mach number ($30 < \mathcal{M} < 50$), the cooling processes would be efficient preferentially in large clouds with dimensions larger than the Field length scale, $l$, while small clouds with dimensions $< l$ are likely to be ablated by the thermal conduction in few $\tau_{\rm cc}$.

## 4. Discussion and concluding remarks

We studied the hydrodynamics of the interaction of an isolated dense cloud with an interstellar shock-wave of an evolved SNR shell, taking into account the effects of both the radiative cooling and the thermal conduction. We explored two complementary cases in which one or the other of these physical processes plays a dominant role in the dynamics. In addition, the effects of the thermal conduction and of the radiative losses were identified by comparing models calculated with these physical processes turned either on or off.

To study the pure hydrodynamic evolution with high accuracy, we considered adiabatic 3-D simulations without thermal conduction and radiative losses; the spatial resolution of these simulations is the highest ever obtained in 3-D shock-cloud interaction simulations. Such a resolution allowed us to describe appropriately the hydrodynamic instabilities developing at the cloud boundaries (they are resolved down to 0.0076 pc in HY2) and, therefore, to evaluate accurately the mass exchange between the cloud and the ambient medium. According to previous results (e.g. KMC94), we found that the SNR shock triggers the development of hydrodynamic instabilities at the cloud boundaries, which destroy the cloud after several $\tau_{\rm cc}$. In this case, the cloud mass loss rate is very efficient, i.e. $\sim 50\%$ of the initial cloud mass is mixed with the ambient medium at $t \approx 3\,\tau_{\rm cc}$.

We then compared the above models to other models accounting for both the thermal conduction and the radiative losses. Since the hydrodynamic instabilities are efficiently suppressed by the thermal conduction, the evolution can be accurately described with 2-D simulations. The thermal conduction plays a dominant role in the evolution of moderately over-dense parsec-size clouds crushed by a 50 Mach shock (post-shock temperature $\approx 4.7 \times 10^6$ K), since $[\tau_{\rm cond}]_{\rm psh} < [\tau_{\rm rad}]_{\rm cl}$ for structures smaller than $\sim 0.8$ pc (see Appendix B), i.e. over a distance comparable to the size of the crushed cloud (see also the location of HTR50 in Fig. 2). The main effect of the thermal conduction is to smooth out the temperature and density inhomogeneities. The hydrodynamic instabilities responsible of the cloud destruction are therefore strongly suppressed, and the cloud becomes more stable and survives for a longer time. At the cloud boundary, the temperature and density gradients (very steep in the model without thermal conduction) are reduced, building up a broad transition region from the inner portion of the cloud to the ambient medium. During evolution, the cloud expands and gradually evaporates. The cloud does not fragment into cloudlets. The mass loss of the cloud is driven by the thermal conduction and is less efficient than in the presence of hydrodynamic instabilities.

The radiative losses play a crucial role for the $\mathcal{M} = 30$ case (post-shock temperature $\approx 1.6 \times 10^6$ K) since, for the conditions of the shocked cloud gas, $[\tau_{\rm rad}]_{\rm cl} < [\tau_{\rm cond}]_{\rm psh}$ for structures larger than $\approx 0.3$ pc (see Appendix B and the location of HTR30 in Fig. 2). The different structure of the shock transmitted into the cloud leads to the formation of dense and cold gas there. On the other hand, Eq. 12 suggests that the thermal conduction is effective in suppressing hydrodynamic instabilities on sub-parsec scales. The shocked cloud evolves into a dense and cold core – unaffected by heat conduction – surrounded by a hot and diluted corona, where the conducted heat exceeds the cooling. The core ultimately fragments into dense, cold and compact filaments, consistent with previous works (Mellema et al. 2002; Fragile et al. 2004). The corona, instead, expands, and evaporates under the effect of the thermal conduction. This is the main mechanism of mass loss of the cloud in this case. The complete evaporation of the corona leaves a "naked" fragmented core collapsing under the effect of the radiation. The cloud keeps its identity as long as the corona surrounds the whole fragmented core.

Note that some details of the simulations depend on the choice of the parameters. For instance, the formation of the dense and cold core or the evaporation of the whole cloud depends also on the density and dimensions of the cloud. However, the two complementary cases $\mathcal{M} = 30$ and $\mathcal{M} = 50$ that we present here are representative of regimes in which either the radiative losses or the thermal conduction play a dominant role.

The results presented here are important for the study of middle-aged X-ray SNR shells whose morphology is affected by ISM inhomogeneities. The examples include the Cygnus Loop (e.g. Patnaude et al. 2002 on the detection of an isolated ISM cloud in the South-East part of the shell), the Vela SNR (e.g. Miceli et al. 2005, on the XMM-Newton observation of an ISM feature in the northern part of the remnant) and G272.2-3.2 (e.g. Egger et al. 1996; on the multi-wavelength observation of an ISM cloud hit by the shock). In other cases, the unfavorable location of the system in the $\mathcal{M} - \chi$ plane (Fig. 2) or the cloud destruction by SN progenitor wind may lead to difficulties in the detection of observable effects of shock interactions with clouds.

Future steps in this project include: the study of the deviations from equilibrium ionization occurring during the complex shock-cloud interaction; the synthesis, from the numerical simulations, of spatially and spectrally resolved X-ray observations with the latest instruments (e.g. Chandra, XMM-Newton, Astro-E2), amenable to direct comparison with SNR observations made with the same instruments; proper account for an ambient magnetic field along with its effect on thermal conduction and on radiative losses.

## Appendix A: Spitzer's thermal conduction in the FLASH code

The thermal conduction is added to the FLASH code using an operator-splitting method with respect to the hydrodynamic evolution: the heat losses (or gains) due to the thermal conduction are added as a source term to the energy equation. The heat flux in the formulation of Spitzer (1962) is calculated by explicit finite difference as:

$$F_i^{(\text{spi})} = -\frac{\kappa_i + \kappa_{i-1}}{2} \times \frac{T_i - T_{i-1}}{\Delta x}, \quad (A.1)$$

where $\kappa_i$ and $T_i$ are the Spitzer's thermal conductivity and the plasma temperature, respectively, at the $i$-th grid point and $\Delta x$ is the cell size. Analogously, the saturated heat flux is:

$$F_i^{(\text{sat})} = -\text{sign}\left(\frac{T_i - T_{i-1}}{\Delta x}\right) \times$$
$$\times 5\phi \left(\frac{P_i + P_{i-1}}{2}\right)^{3/2} \left(\frac{\rho_i + \rho_{i-1}}{2}\right)^{-1/2}, \quad (A.2)$$

where $P_i$ and $\rho_i$ are the plasma pressure and mass density, respectively, at the $i$-th grid point. The total heat flux, including saturation effects, is:

$$F_i = \left(\frac{1}{F_i^{(\text{spi})}} + \frac{1}{F_i^{(\text{sat})}}\right)^{-1}. \quad (A.3)$$

The heat flux is then added to the energy flux generated by the PPM module. This addition is done before any of the zones are updated in the hydrodynamic step. This grants conservation, since the total flux (including the thermal flux) will be corrected during the flux conservation step.

Since the thermal conduction is solved explicitly, a time-step limiter is required to avoid numerical instability. Stability is guaranteed for $\Delta t < 0.5 \, \Delta x^2/D$, where $D$ is the diffusion coefficient, related to the conductivity, $\kappa$, and to the specific heat at constant volume, $c_v$, by $D = \kappa/(\rho c_v)$.

We have verified the FLASH code conduction module using test problems with known analytic solutions. The test case we considered is the propagation of a plane conduction front in a uniform, high temperature plasma, with negligible saturation effects (cf. Reale 1995). Since the test includes the plasma hydrodynamics, the propagation of the conduction front is slightly complicated by the presence of the plasma dynamics. The presence of a thermal front causes also a strong pressure wave which eventually drives significant plasma motion in the same direction as the conduction front. However, as we show below, the mean propagation speed of the conduction is much higher than the mean plasma sound speed and the much faster front can be considered propagating as a pure conduction front.

In the case of a plane, pure conduction, front an analytic solution is available as a self-similar solution (Zel'dovich & Raizer 1966). Defining the dimensionless parameter

$$\xi = \frac{x}{(\kappa Q^n t)^{1/(n+2)}} \quad (A.4)$$

where $Q$ is the integral of $T$ over the whole space, $n = 5/2$, and $\kappa = 9.2 \times 10^{-7}$ (typical of coronal plasma) in our case, the solution is given by

$$T = T_c \left(1 - \frac{x^2}{x_f^2}\right)^{1/n}, \quad (A.5)$$

where

$$T_c = \left(\frac{Q^2}{\kappa t}\right)^{1/(n+2)} \left(\frac{n}{2(n+2)}\xi_0^2\right)^{1/n}, \quad (A.6)$$

$$x_f = (\kappa Q^n t)^{1/(n+2)} \xi_0, \quad (A.7)$$

$$\xi_0 = \left[\frac{(n+2)^{1+n} \, 2^{1-n}}{n\pi^{n/2}} \left(\frac{\Gamma(1/2 + 1/n)}{\Gamma(1/n)}\right)^n\right]^{1/(n+2)} \quad (A.8)$$

and $\Gamma$ is the gamma function.

In our test, we took as the initial condition the analytic solution at $t = 0.1$ s for a plasma with a particle density of $10^9$ cm$^{-3}$ and $Q = 1.2 \times 10^{15}$ K cm, so that the maximum initial temperature is at $3.65 \times 10^6$ K. Reflecting boundary conditions have been assumed at the left boundary and outflow boundary conditions (zero-gradient) at the right boundary.

In our test simulations, the conduction front propagates through $\sim 2 \times 10^8$ cm in 3 s with a mean propagation speed $\sim 700$ km/s, i.e. much larger than the mean plasma sound speed (for $T \sim 2 \times 10^6$ K, $c_s \sim 200$ km/s). The front, therefore, propagates almost as a pure conduction front and we can compare the numerical solution with the analytic solution (Eq. A.5). Fig. A.1 compares the temperature distributions computed numerically and analytically, and shows a good agreement between them. We expect a departure from the analytic solution as soon as the conduction speed (which is $\propto T^{5/2}$) is significantly reduced at lower temperatures and approaches the plasma local sound speed ($\propto T^{1/2}$), as it happens at later times when the front temperature is reduced significantly from the initial high value. We checked that the local plasma bulk velocity increases with time, although remaining well below the sound speed; also the density begins to change significantly only at late times, when a density front is forming, as a consequence of the pressure front.

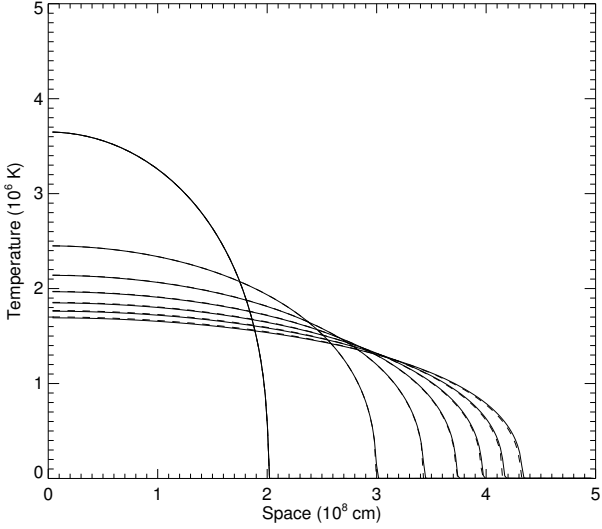

**Figure A.1.** Comparison of the temperature distributions along the direction of propagation as computed numerically with the FLASH code (solid lines), sampled every 0.5 s, with the corresponding analytic solutions (dashed lines).

## Appendix B: Time-scales in terms of $\mathcal{M}$ and $\chi$

The effect of the thermal conduction on the dynamics of the shock-cloud interaction is evaluated by comparing the time-scale for conduction in the shocked ambient medium with the cloud crushing time. In particular, using Eqs. 8 and 10, we derive the condition:

$$\frac{[\tau_{\rm cond}]_{\rm psh}}{\tau_{\rm cc}} = 2.6 \times 10^{-9} \, \frac{n_{\rm psh} l^2}{T_{\rm psh}^{5/2}} \, \frac{\beta^{1/2} w}{\chi^{1/2} r_{\rm cl}} < 1 \; . \quad (B.1)$$

We use Eqs. 6 and 7 to express $T_{\rm psh}$ and $n_{\rm psh}$ as

$$T_{\rm psh} = \left(\frac{2}{\gamma+1}\right)^2 \frac{\gamma-1}{2} \frac{\mu m_H}{2k_{\rm B}} \mathcal{M}^2 c_{\rm ism}^2 \quad (B.2)$$

$$n_{\rm psh} = 4 \, n_{\rm ism}$$

Substituting Eqs. B.2 into Eq. B.1

$$\frac{[\tau_{\rm cond}]_{\rm psh}}{\tau_{\rm cc}} = 7.8 \times 10^{-3} \beta^{1/2} \, \frac{n_{\rm ism}}{T_{\rm ism}^2} \, \frac{l^2}{\mathcal{M}^4 \chi^{1/2} r_{\rm cl}} < 1 \; . \quad (B.3)$$

The radiative losses influence the shocked cloud dynamics when the cooling time-scale behind the shock transmitted into the cloud is short compared with the cloud-crushing time. Using Eqs. 8 and 11, the above condition is expressed as:

$$\frac{[\tau_{\rm rad}]_{\rm cl}}{\tau_{\rm cc}} = 2.5 \times 10^3 \, \frac{T_{\rm scl}^{3/2}}{n_{\rm scl}} \, \frac{\beta^{1/2} w}{\chi^{1/2} r_{\rm cl}} < 1 \quad (B.4)$$

where $T_{\rm scl}$ and $n_{\rm scl}$ are the temperature and the density of the shocked cloud, respectively. We use the relation (e.g. KMC94; McKee & Cowie 1975)

$$w_{\rm cl}^2 = \frac{\beta w^2}{\chi} \quad (B.5)$$

and Eqs. 6 and 7 to express $T_{\rm scl}$ and $n_{\rm scl}$ as:

$$T_{\rm scl} = \frac{\beta T_{\rm psh}}{\chi} \quad (B.6)$$

$$n_{\rm scl} = 4 \, n_{\rm cl} = 4 \, \chi \, n_{\rm ism}$$

Substituting Eqs. B.6 into Eq. B.4 and using Eqs. B.2

$$\frac{[\tau_{\rm rad}]_{\rm cl}}{\tau_{\rm cc}} = 5.8 \times 10^5 \beta^2 \, \frac{T_{\rm ism}^2}{n_{\rm ism}} \, \frac{\mathcal{M}^4}{r_{\rm cl} \chi^3} < 1 \; . \quad (B.7)$$

Finally, we compare the cooling time-scale behind the shock transmitted into the cloud with the thermal conduction time-scale in the shocked ambient medium. From Eqs. B.3 and B.7, we derive

$$\left(\frac{[\tau_{\rm rad}]_{\rm cl}}{[\tau_{\rm cond}]_{\rm psh}}\right)^{1/2} = 8.6 \times 10^3 \beta^{3/4} \, \frac{T_{\rm ism}^2}{n_{\rm ism}} \, \frac{\mathcal{M}^4}{\chi^{5/4} l} > 1 \quad (B.8)$$

McKee & Cowie (1975) derived a detailed expression for the numerical factor $\beta$:

$$\beta = \beta' \frac{\gamma_{\rm cl} + 1}{\gamma + 1} F \quad (B.9)$$

where $\quad F \approx 3.15 - 4.78 \frac{w_{\rm cl}}{w} + 2.63 \left(\frac{w_{\rm cl}}{w}\right)^2$

the numerical factor $\beta'$ is 1 at the shock and decreases behind the shock and $\gamma_{\rm cl}$ is the ratio of specific heat in the cloud material. Setting $\beta' = 1$ and $\gamma_{\rm cl} = \gamma$, the above expression reduces to $\beta = F$. Substituting $F$ into Eq. B.5

$$(2.63 - \chi)x^2 - 4.78x + 3.15 = 0 \quad (B.10)$$

where $x = w_{\rm cl}/w$. The above equation has the solution

$$x = \frac{2.39 - 1.8\sqrt{\chi - 0.8}}{2.63 - \chi} \quad (B.11)$$

leading to

$$\beta = \chi \left(\frac{w_{\rm cl}}{w}\right)^2 = \chi \left(\frac{2.39 - 1.8\sqrt{\chi - 0.8}}{2.63 - \chi}\right)^2 \quad (B.12)$$

which ranges between 1 and 2.5 for $3 < \chi < 100$.

For the cases considered in this paper, $\chi = 10$ and $\beta \approx 1.7$. From Eq. B.3, the thermal conduction time-scale is shorter than the cloud crushing time on scales below $l \approx 0.8$ pc in the $\mathcal{M} = 50$ shock case and $l \approx 0.3$ pc in the $\mathcal{M} = 30$ case: thermal gradients smaller than that will be diffused on time-scales shorter than $\tau_{\rm cc}$. These numbers suggest that hydrodynamic instabilities, which in our problem develop on sub-parsec scales and on time-scales shorter than the cloud crushing time, are suppressed by thermal conduction in both the cases considered. In addition, our estimate suggests that, in the $\mathcal{M} = 50$ shock case, the cloud itself, has a radius comparable to the characteristic length-scale $l$ and, therefore, is likely to "evaporate" due to the thermal conduction on time-scale of the order of $\tau_{\rm cc}$.

From Eq. B.7, in the $\mathcal{M} = 50$ shock case, $[\tau_{\rm rad}]_{\rm cl}/\tau_{\rm cc} \approx 3.4$, indicating that the shock transmitted into the cloud is regulated by the energy losses on time-scales of the order of $3\,\tau_{\rm cc}$, i.e. the time-scale on which we focused on our simulations. On the other hand, in the $\mathcal{M} = 30$ shock case $[\tau_{\rm rad}]_{\rm cl}/\tau_{\rm cc} \approx 0.4$, indicating that the transmitted shock is strongly radiative and its evolution is regulated by the energy losses on time scales shorter than the cloud crushing time.

*Acknowledgements.* SO acknowledges the hospitality of the Flash Center at the University of Chicago, where part of the present work was carried out. We thank the anonymous referee for his/her useful comments on the manuscript. This work is supported in part by the U.S. Department of Energy under Grant No. B523820 to the Center of Astrophysical Thermonuclear Flashes at the University of Chicago (USA). The software used in this work was in part developed by the Center of Astrophysical Themonuclear Flashes. The 3-D simulations were performed on the IBM/Sp4 machine at CINECA (Bologna, Italy) and the 2-D simulations on the Compaq cluster at the SCAN (Sistema di Calcolo per l'Astrofisica Numerica) facility of the INAF - Osservatorio Astronomico di Palermo. This work was supported in part by Ministero dell'Istruzione, dell'Università e della Ricerca and by Istituto Nazionale di Astrofisica.